# Sunspot Observations Made by Hallaschka during the Dalton Minimum

V.M.S. Carrasco[1,2] (ORCID: 0000-0001-9358-1219), J.M. Vaquero[2,3] (ORCID: 0000-0002-8754-1509), R. Arlt[4], M.C. Gallego[1,2] (ORCID: 0000-0002-8591-0382)

[1] Departamento de Física, Universidad de Extremadura, 06071 Badajoz, Spain [e-mail: vmscarrasco@unex.es]

[2] Instituto Universitario de Investigación del Agua, Cambio Climático y Sostenibilidad (IACYS), Universidad de Extremadura, 06006 Badajoz, Spain

[3] Departamento de Física, Universidad de Extremadura, 06800 Mérida, Spain

[4] Leibniz Institute for Astrophysics Potsdam, An der Sternwarte 16, 14482 Potsdam, Germany

**Abstract:** We present and analyse the sunspot observations performed by Franz I. C. Hallaschka in 1814 and 1816. These solar observations were carried out during the so-called Dalton minimum, around the maximum phase of the Solar Cycle 6. These records are very valuable because they allow us to complete observational gaps in the collection of sunspot group numbers, improving its coverage for this epoch. We have analysed and compared the observations made by Hallaschka with the records made by other contemporary observers. Unfortunately, the analysis of the sunspot areas and positions showed that they are too inaccurate for scientific use. But, we conclude that sunspot counts made by Hallaschka are similar to those made by other astronomers of that time. The observations by Hallaschka confirm a low level of the solar activity during the Dalton minimum.

## 1. Introduction

The amount of sunspots appearing on the solar disc has a cyclic behaviour (Hathaway, 2015). Approximately, each eleven years a maximum in the number of sunspots can be observed. The two most common indices used to describe this behaviour are the international sunspot number and the group sunspot number (Hoyt and Schatten, 1998; Clette *et al*., 2014). However, it has been proven that these indices present some problems and remarkable discrepancies in some periods, especially in the historical part (Clette *et al*., 2014). Thus, new series related with the sunspot number have been recently published in order to solve the problems detected (Clette and Lefèvre, 2016; Lockwood *et al*., 2016; Svalgaard and Schatten, 2016; Usoskin *et al*., 2016; Chatzistergos *et al*., 2017).

A complete database of sunspot counts is necessary in order to obtain a good sunspot number series using an adequate methodology. In the 19th century, Rudolf Wolf carried out a great compilation of historical sunspot records to establish the relative sunspot number (Wolf, 1856). Hoyt and Schatten (1998) increased the number of sunspot records beyond the series of Wolf by reducing the count to the number of sunspot groups, not including the number of single sunspots. Recently, Vaquero *et al*. (2016) started a survey to obtain a revised collection of sunspot group numbers based on the previous works of Wolf and Hoyt and Schatten (1998). However, improving the database of sunspot records continues and more accurate solar activity indices will be obtained. For example, there are observational gaps in the first part of the telescopic era for these databases, including the periods of low solar activity, the so-called Maunder (1645-1715) and Dalton minima (1793-1827).

The grand minimum epochs or periods with a very reduced solar activity are of great interest for the scientific community. They are key periods to understand the long-term solar activity behaviour and its influence on the heliosphere and our planet (Haigh, 2007; Usoskin, 2017). Dalton minimum was the period between the end of the 18th century and the first quarter of the 19th century characterized by a reduced solar activity. However, the Dalton minimum is not considered as a grand minimum of solar activity like, for example, the Maunder minimum (Usoskin, Solanki, and Kovaltsov, 2007). Considering the Solar Cycles 5 and 6 which correspond approximately to the Dalton minimum, the historical records cover less than 50 % of the total number of days for those periods although, fortunately, the solar observations are uniformly distributed in time (Vaquero *et al.*, 2016). Therefore, the incorporation to the database of group counts of new information about sunspot records not previously published is important to improve the temporary coverage of this period. Thus, the recovery of a sunspot drawing made by Iwahashi Zenbei (Hayakawa *et al.*, 2018) on 26 August 1793 and an undated one, during the decline phase of the Solar Cycle 4, and a set of 25 drawings by Jonathan Fisher (Denig and McVaugh, 2017) in 1816 and 1817, in the maximum of the Solar Cycle 6, have helped to improve the characterization of the solar activity corresponding to the Dalton minimum.

We found sunspot observations performed by Cassian Hallaschka in 1814 and 1816 (shortly before the maximum of the Solar Cycle 6) which have not yet been included in sunspot studies. The aim of this work is to present these solar observations and to contribute to improve the reconstruction of solar activity of the past, in particular, of the Dalton minimum. In Section 2, we indicate some biographical notes about Hallaschka and also describe his observations. An analysis of these data is shown in Section 3. In that section, we also compare the sunspot observations made by Hallaschka with other sunspot observations of the same time. Section 4 is devoted to present the main conclusion of this work.

## 2. Biographical Notes and Observational Data

Franz Ignatz Cassian Hallaschka (1780–1847) was a Czech physicist, born in the Moravia region, who performed meteorological and astronomical observations during the first years of the 19th century (Brázdil *et al*., 2016). In 1807, he obtained the PhD degree at the University of Vienna and he returned to the Czech Republic to work as a professor of mathematics and physics in Mikulov and Brno until 1814 (Šolc, 1999). In that year, he moved to Prague to be professor of physics at the University of Prague, where he was elected as the Chancellor of the University in 1832. Moreover, he was in contact with important European scientists as, for example, Friedrich Wilhelm Bessel. Hallaschka also carried out systematic meteorological observations of, for example, atmospheric pressure and temperature. Among its astronomical publications, along with sunspot records, we highlight his book *Elementa eclipsium quas patitur tellus, luna eam inter et solem versante ab A. 1816 usque ad A. 1860* where Hallaschka calculated all solar eclipses from 1816 to 1860 and showed maps including the trajectories for those eclipses.

Among these studies, we have recovered sunspot observations made by Hallaschka (Hallaschka, 1814) at Brno (around 49°11′ N, 16°36′ E), where he established a small astronomical observatory, from 9 April to 3 May 1814. Furthermore, two years later, when Hallaschka was professor at the University of Prague (around 50°05′ N, 14°25′ E), again reported sunspot observations (Hallaschka, 1816a; 1816b). In that moment, he lived in the Convent of the Piarists in the New Town of Prague and adapted the top floor of this building as observatory. This observatory had good equipment with several achromatic telescopes, a 0.2-meter (8 inches) multiplication theodolite, a mirror sextant 0.25 meters (10 inches) in radius, a pendulum, and even barometers, thermometers, and magnetic needles, *inter alia* (Šolc, 1999).

We know that Hallaschka used a 0.69-meter (27 inch) focal-length Ramsden telescope to carry out the sunspot observations in both 1814 and 1816. This information appears in the descriptions of his observations on 9 April 1814 and 28 February 1816. However, we also know that Hallaschka employed additional telescopes to improve his sunspot observations. In 1814, Hallaschka reported that he observed with a Gregorian reflecting telescope of 100 mm (4 inches) in aperture and nearly 84 times magnification on 9 April 1814 and with a 1.47-meter (58 inch) telescope with 105 mm (41 Parisian lines) aperture on 19 April 1814. Hallaschka indicated that he was able to perform a more accurate observation with these two telescopes. Note that one Parisian line is equal to 2.558 mm (Cardarelli, 2003). On 15 March 1816, he also employed a new telescope to improve the quality of the observation, a Fraunhofer achromatic telescope with a 1.47 meter (58 inch) focal length. Although we do not know exactly what telescopes were used by Hallaschka for each observing day, Hallaschka registered in the description of the sunspot observation corresponding to 15 March 1816 that all previous observations were performed with the Ramsden telescope. Therefore, it seems that Hallaschka generally used the Ramsden telescope for his solar observations but also employed other telescopes when he wanted to be more accurate in the observations. Unfortunately, we do not know exactly the method employed by Hallaschka to observe the sunspots. According to the exaggerated sizes and the probably unreliable positions of the sunspots recorded in his drawings (see Section 3), we suppose that Hallaschka observed through the telescope to draw the images approximately, rather than using the projection method for direct drawings. In his annotations, Hallaschka also mentions that he used a sextant to measure the diameters of sunspots (Hallaschka, 1816a).

The sunspot observations made by Hallaschka recovered in this work were published in *Hesperus*, a journal of general information written by Christian Carl Andre in German. In addition to these sunspot records, Hallaschka also published his meteorological observations in that journal. In *Hesperus*, Hallaschka shows a description of his sunspot observations reporting information as dates and time of the observations, sunspot group counts (sometimes single sunspot counts), evolution of the sunspots on the solar disc according to their location, size and form, and the telescopes used for the observations as mentioned above. Moreover, we highlight that we have found nine sunspot drawings made by Hallaschka in this documentary source (Figure 1). A digital version of these drawings can be consulted at: https://www.digitale-sammlungen.de/. We also note that only other three sunspot observations, not discussed here, made by Hallaschka (24-26 June 1819) are included in the database of Hoyt and Schatten (1998) and Vaquero *et al*. (2016). Therefore, the records presented in this work have not been taken into account for the recent reconstructions of the solar activity. Furthermore, using the data recovered in this article, 10 and 9 days without any record in the databases can be covered now for 1814 and 1816 respectively. Without these new recovered observations made by Hallaschka, the temporary coverage of the sunspot records in Vaquero *et al*. (2016) for 1814 is equal to 51.0 % and 67.1 % for 1816. These values mean that we have not a great temporary coverage for this period (in particular for 1814) and it became very important to recover new observations for a better characterization of the solar activity. After the incorporation of these recovered data by Hallaschka, the temporary coverages for 1814 and 1816 increase to 53.6% and 69.4 % respectively.

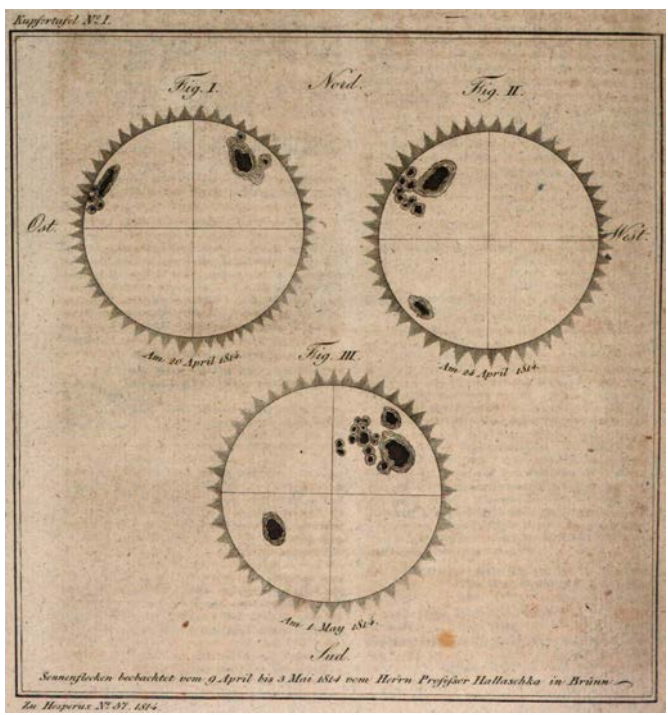

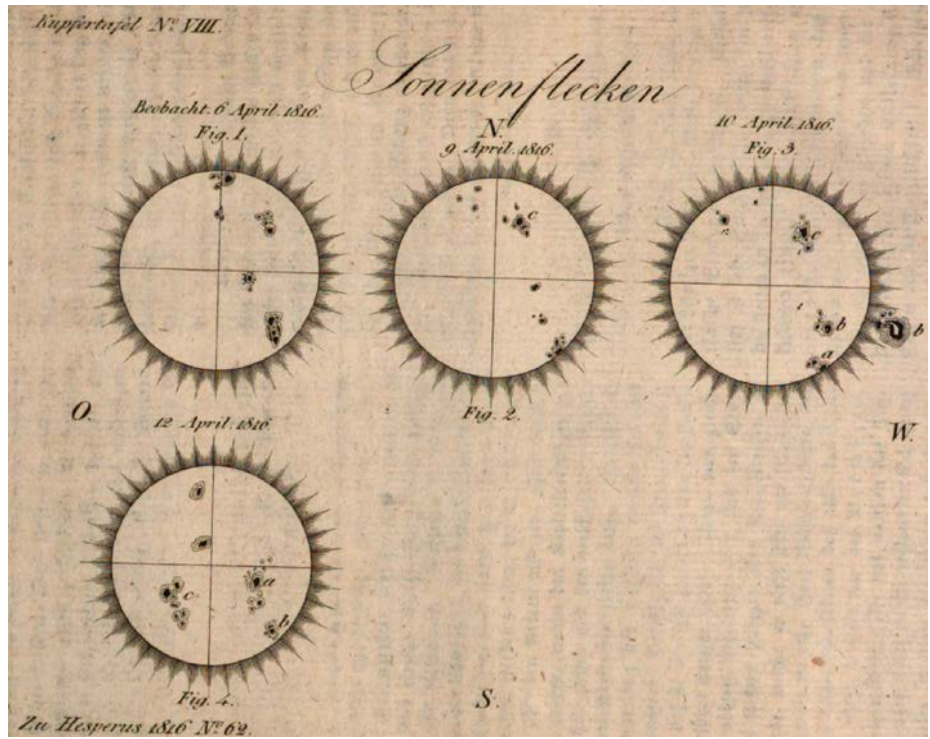

Figure 1. Sunspot drawings made by Hallaschka on: (top-left panel) 20 and 24 April and 1 May 1814, (top-right panel) 10 and 15 March 1816 and (bottom panel) 6, 9, 10, and 12 April 1816 [Source: Hallaschka (1814, 1816a, 1816b)].

**3. Results and Discussion**

We have recovered the sunspot observations made by Hallaschka published in the journal *Hesperus* in 1814 and 1816. Table 1 lists the dates and group numbers recorded by Hallaschka. For 1814, we have retrieved 16 records from 9 April to 3 May, while for 1816, we recovered 30 sunspot observations made during the first eight months of that year. The record number made by Hallaschka is not very numerous (46) but they clearly improve the temporal coverage of the sunspot group counts in the Dalton minimum. We highlight that Hallaschka reported the observation of a new sunspot group on 13 July 1816 and the other one on 16 July 1816 but he does not mention if the one sunspot group present on the solar disc from 6 July 1816 was still visible on 13 and 16 July. For this reason, Table 1 shows sunspot group counts equal to 1-2 on 13 July and 2-3 on 16 July. Furthermore, the gaps in the records corresponding to 1816 were caused by the bad meteorological conditions for the observations. Hallaschka reported that the sky was mostly cloudy from 15 March to 4 April and from the second part of April to June. Hallaschka also indicated that the second half of June was rainy and he could not observe from August due to bad meteorological conditions. Note that that year, 1816, is commonly known as "the year without a summer" due to cold temperatures which affected the northern hemisphere (Harington, 1992; Trigo *et al*., 2009; Luterbacher and Pfister, 2015).

Table 1. Group numbers registered by Hallaschka during the years 1814 and 1816.

| DATE | GROUPS | DATE | GROUPS | DATE | GROUPS |
|---|---|---|---|---|---|
| 1814/4/9 | 1 | 1816/2/28 | 2 | 1816/7/1 | 0 |
| 1814/4/10 | 1 | 1816/2/29 | 2 | 1816/7/3 | 1 |
| 1814/4/11 | 2 | 1816/3/1 | 2 | 1816/7/6 | 1 |
| 1814/4/12 | 2 | 1816/3/2 | 2 | 1816/7/7 | 1 |
| 1814/4/13 | 2 | 1816/3/4 | 2 | 1816/7/8 | 1 |
| 1814/4/14 | 1 | 1816/3/6 | 1 | 1816/7/9 | 1 |
| 1814/4/16 | 1 | 1816/3/9 | 0 | 1816/7/10 | 1 |
| 1814/4/17 | 1 | 1816/3/10 | 4 | 1816/7/11 | 1 |
| 1814/4/19 | 1 | 1816/3/15 | 4 | 1816/7/13 | 1 - 2 |
| 1814/4/20 | 2 | 1816/4/4 | 2 | 1816/7/16 | 2 - 3 |
| 1814/4/22 | 2 | 1816/4/5 | 2 | 1816/7/24 | 0 |
| 1814/4/24 | 2 | 1816/4/6 | 5 | 1816/7/25 | 0 |
| 1814/4/29 | 2 | 1816/4/9 | 7 | 1816/7/26 | 0 |
| 1814/5/1 | 2 | 1816/4/10 | 6 | 1816/8/1 | 2 |
| 1814/5/2 | 2 | 1816/4/12 | 5 | | |
| 1814/5/3 | 1 | 1816/6/19 | 0 | | |

The importance of these observations is that they were carried out during Solar Cycle 6 in the time known as the Dalton minimum. As was aforementioned, the Dalton minimum was a period of a reduced solar activity level which occurred between 1793 and 1827, approximately. Figure 2 depicts the sunspot number index (version 2, http://www.sidc.be/silso/datafiles) considering the Solar Cycles from 4 to 7 (from 1784 to 1833), *i.e*., during the Dalton minimum. In Figure 2, it can be seen that the maximum for the Solar Cycle 6 lies in 1816, coinciding with one of the years in which Hallaschka observed. Therefore, the observations corresponding to 1814 lie in the ascent stage of the solar cycle.

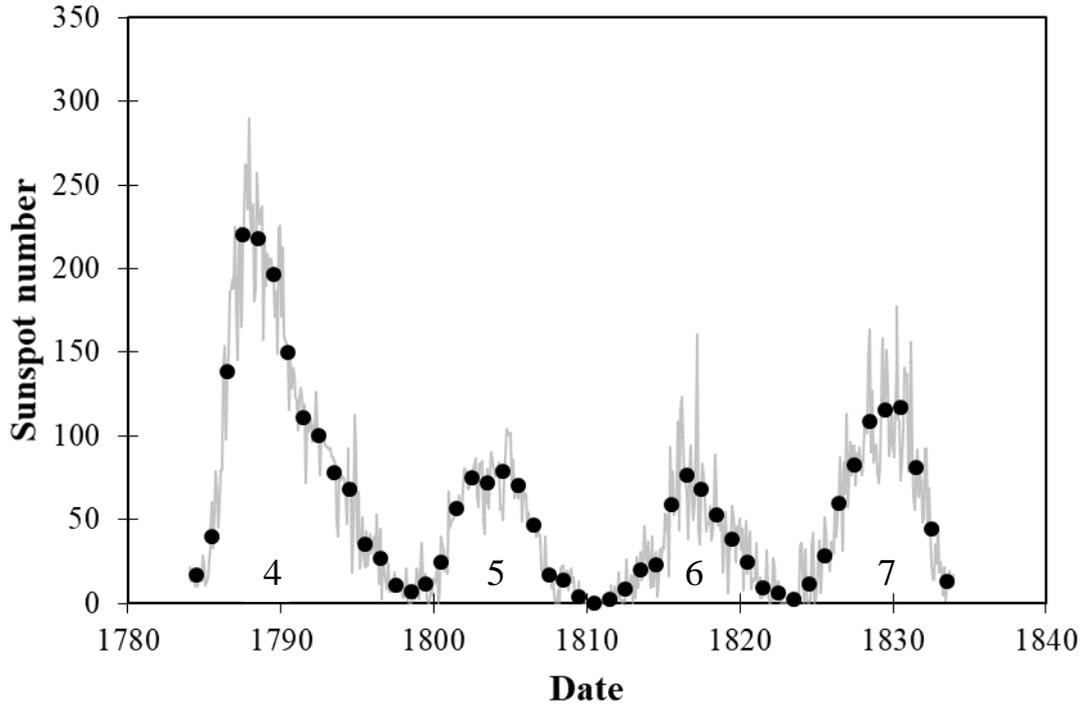

Figure 2. Monthly (grey line) and yearly (black dots) sunspot number index version 2 from Solar Cycle 4 to Solar Cycle 7.

The day with a greater level of solar activity registered by Hallaschka was the 9th April 1816 when Hallaschka observed 7 sunspot groups (Figure 1). From Vaquero *et al*. (2016), it is verified that this is the second-largest group count for 1816 after 8 sunspot groups observed by Tevel on 26 May 1816. However, the greater count of single sunspots reported by Hallaschka was on 12 April 1816 when he himself counted 46 sunspots distributed in 5 groups. In the database of Hoyt and Schatten (1998) and Vaquero *et al*. (2016), it can be seen that there is another sunspot observation made by Lindener on 9 April 1816. Lindener, as Hallaschka, also registered 7 sunspot groups for that same day (Figure 3). Figure 3 contains information about the raw sunspot **group** counts made by all observers included in Vaquero *et al*. (2016) for the period 1813–1818 along with the sunspot observations made by Hallaschka, studied in this work, and Fisher, recently published by Denig and McVaugh (2017). For the remaining observations, the number of sunspot groups observed by Hallaschka is similar than those observed by other astronomers (Figure 3). The observations made by Fisher, recovered by Denig and McVaugh (2017), are composed by 25 sunspot drawings. Unfortunatelly, there are not sunspot records by Hallaschka and Fisher for the same day to directly compare both sets of observations. The closest observation dates are: i)

Hallaschka observed 1 sunspot group on 11 and between 1 or 2 sunspot groups on 13 July 1816 (as explained above, Hallaschka registered a new sunspot group on 13 July but he does not mention if the one sunspot group present on the solar disc from 6 to 11 July 1816 was still visible on 13 July) and Fisher indicated 2 groups on the intermediate day, 12 July 1816; and ii) Hallaschka registered 2 sunspot groups on 1 August 1816 while Fisher just 1 on 30 July 1816. It is worth mentioning that according to Hallaschka's annotations, he observed 2 sunspot groups near the centre of the solar disc on 1 August 1816. These groups must therefore have been close to each other and they could have been considered as the same group by Fisher. Thus, these data indicate the similarity between the records of these two new-recovered observers according to the sunspot group numbers registered by both.

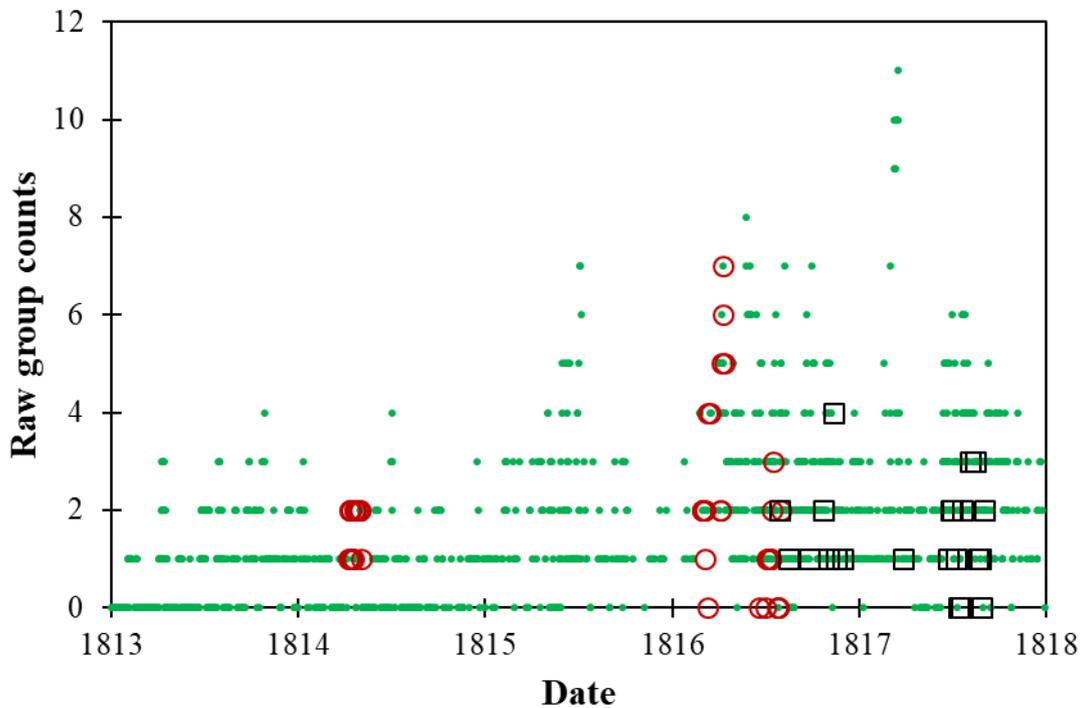

Figure 3. Raw group counts according to: i) revised collection of group numbers by Vaquero *et al.* (2016) for the period 1813–1818 (green dots), ii) sunspot observation by Hallaschka recovered in this work (red circles), and records by Fisher published by Denig and McVaugh (2017) (black squares).

The drawings by Hallaschka contain a well-defined coordinate system with compass directions indicated. Since the telescopes were very likely sitting on equatorial mounts, we may assume that the coordinate system given is a celestial one, *i.e.* the horizontal line is a parallel to the celestial equator. Using the solar ephemeris from the JPL Horizons service (https://ssd.jpl.nasa.gov/horizons.cgi), we can convert the Cartesian coordinates measured in the drawings into heliographic coordinates. As one can guess already by eye, many spot latitudes are far away from the solar equator. Many spots have latitudes of beyond ±40º, with the extreme of -60º for the southern spot on 1814 April 24. Hallaschka's knowledge of the various coordinate systems must have been rather limited, since he mentioned several times that the spots crossed the solar equator, according to his observations.

We tried to rectify the coordinate system by manually setting a position angle of the solar disk, such that the absolute values of the latitudes were minimized. As a result, there are no latitudes beyond ±50º anymore, but still several of ±40º and higher. Given that the observations were made slightly before the maximum of Solar Cycle 6 in 1816/1817 (Hathaway 2015), high latitudes of sunspot groups are not unlikely, but we believe one should not use the positions for scientific purposes. The only observation delivering reasonable results is the one of 10 March 1816. Figure 4 shows the resulting heliographic coordinate grid on the drawing. The tilt angles of all the groups on that day look very similar. Assuming they are near 5-6 degrees as expected from modern data, one may argue that the position angle was even a bit more negative, moving all groups into the northern hemisphere. Tilt angles have a large scatter on the one hand, but large, evolved groups such as the ones shown here tend to have a somewhat smaller scatter (*e.g.* Kosovichev and Stenflo, 2008), in agreement with the drawing of 10 March 1816. Together with the assumption of no too high latitudes for any of the groups, we may at least subjectively guess the solar equator for all the drawings and divide subjectively the groups into northern-hemisphere and southern-hemisphere ones. Table 2 gives these subjective results for the nine drawings by Hallaschka. We can see according to Table 2 that the northern hemisphere was more active than the southern hemisphere when Hallaschka carried out his sunspot observations. Furthermore, the areas of the sunspots are exaggerated. We did not make any attempt in measuring or calibrating the areas of the sunspots. Only a statistical comparison of the area distribution with other area distributions using contemporary data may deliver meaningful results. Unfortunately, the number of spots observed by Hallaschka is too small for this comparison even if a large number of measurements of contemporary sunspot areas were available.

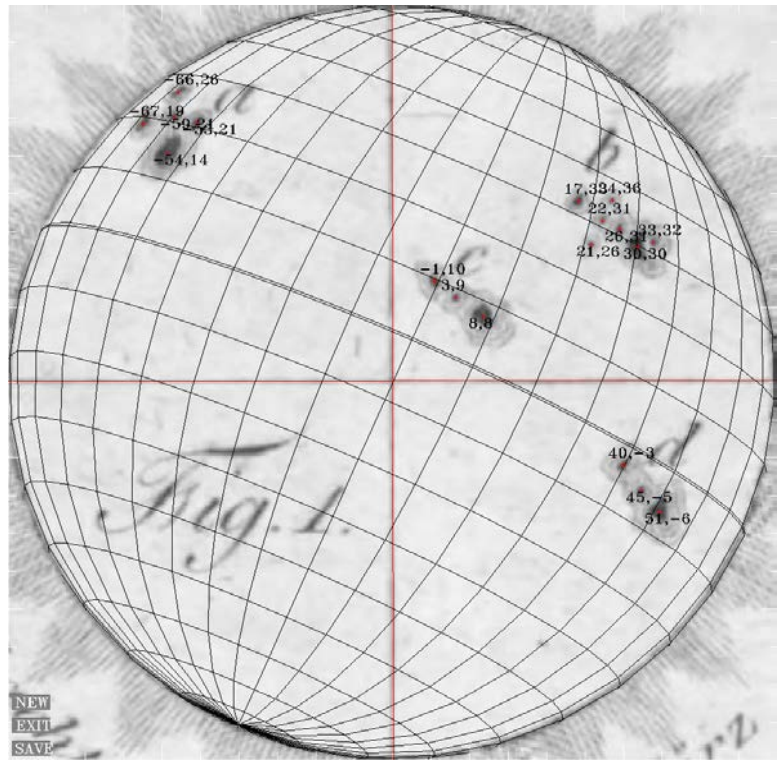

Figure 4. Reconstructed heliographic coordinate grid of 1816 March 10 assuming that the horizontal line is a parallel to the celestial equator. The position angle of the solar axis is -24.0°, the latitude of the solar disk center is -7.2°. The numbers are the central meridian distances and latitudes of the individual spots.

Table 2. Number of northern-hemisphere groups and southern-hemisphere groups for the drawings by Hallaschka.

| Date | Northern hemisphere | Southern hemisphere |
|---|---|---|
| 1814 Apr 20 | 2 | 0 |
| 1814 Apr 24 | 1 | 1 |
| 1814 May 01 | 1 | 1 |
| 1816 Mar 10 | 4 | 0 |
| 1816 Mar 15 | 4 | 0 |
| 1816 Apr 06 | 5 | 0 |
| 1816 Apr 09 | 5 | 2 |
| 1816 Apr 10 | 5 | 1 |
| 1816 Apr 12 | 3 | 2 |

**4. Conclusions**

We present the sunspot observations registered by Hallaschka in 1814 and 1816 during the Dalton minimum. These observations were carried out around the maximum of the Solar Cycle 6. The sunspot records made by Hallaschka analysed in this work are a valuable documentary source because, on the one hand, they were performed in an epoch of reduced solar activity (the Dalton minimum) and, on the other hand, the observation set covers some observational gaps contained in the sunspot group databases of Hoyt and Schatten (1998) and Vaquero *et al.* (2016). We have only been able to recover one month of sunspot records corresponding to 1814 while, for 1816, we have retrieved records that cover half a year. Moreover, in 1816, Hallaschka reports that their periods without observations were due to bad weather (note that 1816 is known as "the year without a summer"). According to his notes, it seems that Hallaschka generally used a 0.69-meter (27 inch) focal length Ramsden telescope to carry out the sunspot observations. However, he sometimes observed with other telescopes (a Gregorian reflecting telescope of 100 mm (4 inches) in aperture, a 1.47-meter (58 inch) telescope with 105 mm (41 Parisian lines, one Parisian line is equal to 2.558 mm) in aperture, and a Fraunhofer achromatic telescope of 1.47 meters (58 inches) focal length) to improve the accuracy of his observations.

We have analysed the sunspot observations made by Hallaschka and compared with sunspot records of that time. Some results on this sense are the following: we have showed that the maximum value according to the group numbers registered by Hallaschka was 7 on 9 April 1816 and, for the same day, it can be seen in Vaquero *et al.* (2016) that Lindener also registered 7 sunspot groups. We have also compared the last sunspot records by Hallaschka in 1816 with the first sunspot observations made by Jonathan Fisher, recently published by Denig and McVaugh (2017), verifying the similarity between these two sets of sunspot observations. Finally, comparing the sunspot observations made by Hallaschka with the sunspot records made by the remaining observers of the same period, we conclude that the sunspot records by Hallaschka agree with the sunspot observations made by astronomers of that time. Furthermore, the sunspot observations made by Hallaschka during the maximum phase of the Solar Cycle 6 agree with a relatively low level of solar activity during the Dalton minimum. Unfortunately, the positional and area information of the observations were found to be too inaccurate for scientific purposes. Finally, we note that the sunspot observations made by Hallaschka should be considered in future reconstructions of the solar activity of the Dalton minimum.

**Acknowledgements**

This research was supported by the Economy and Infrastructure Counselling of the Junta of Extremadura through project IB16127 and grant GR15137 (co-financed by the European Regional Development Fund) and by the Ministerio de Economía y Competitividad of the Spanish Government (AYA2014-57556-P and CGL2017-87917-P). RA acknowledges the support from Deutsche Forschungsgemeinschaft through grant AR355/12-1. The authors have benefited from the participation in the ISSI workshops. Authors acknowledge Rudolf Brazdil for his help in locating the Hallaschka's original sunspot drawings.

**Disclosure of Potential Conflicts of Interest** The authors declare that they have no conflicts of interest.

V.M.S. Carrasco *et al.*